\documentstyle[epsfig]{aa}
\include{psfig}
\begin{document}
\thesaurus{ (04.03.1) (04.19.1) (08.22.1) (11.13.1) }
\title{
Eros variable stars: A catalog of Cepheids in the central regions \\ of the Magellanic Clouds
\thanks{Based on observations made at ESO, La Silla, Chile.}
}
\author{The EROS collaboration \\
C.~Afonso\inst{1}, J.N.~Albert\inst{2}, C.~Alard\inst{3}, A.~Amadon\inst{1},
J.~Andersen\inst{4}, R.~Ansari\inst{2}, {\'E}.~Aubourg\inst{1}, F.~Bauer\inst{1,5}, 
P.~Bareyre\inst{1,5}, J.P.~Beaulieu\inst{3}, G. Blanc\inst{1},
A.~Bouquet\inst{3}, S.~Char\inst{6}$^{\dagger}$, X.~Charlot\inst{1}, 
F.~Couchot\inst{2}, C.~Coutures\inst{1}, F.~Derue\inst{2}, R.~Ferlet\inst{3}, 
C.~Gaucherel\inst{1}, J.F.~Glicenstein\inst{1}, B.~Goldman\inst{1},
A.~Gould\inst{7}\thanks{Alfred P. Sloan Foundation Fellow},
D.~Graff\inst{7,8}, M.~Gros\inst{1}, J.~Haissinski\inst{2},
J.C.~Hamilton\inst{5}, D.~Hardin\inst{1}, J.~de Kat\inst{1}, A.~Kim\inst{5},
T.~Lasserre\inst{1}, {\'E}.~Lesquoy\inst{1}, C.~Loup\inst{3},
C.~Magneville \inst{1}, B.~Mansoux\inst{2}, J.B.~Marquette\inst{3},
{\'E}.~Maurice\inst{9}, A.~Milsztajn \inst{1}, M.~Moniez\inst{2},
N.~Palanque-Delabrouille\inst{1}, O.~Perdereau\inst{2}, L.~Pr{\'e}vot\inst{9},
C.~Renault\inst{1}, N.~Regnault\inst{2}, J.~Rich\inst{1},
M.~Spiro\inst{1}, A.~Vidal--Madjar\inst{3},
L.~Vigroux\inst{1}, S.~Zylberajch\inst{1} 
}
\institute{
     CEA, DSM, DAPNIA, Centre d'{\'E}tudes de Saclay, 91191 Gif-sur-Yvette, Cedex, France
\and Laboratoire de l'Acc{\'e}l{\'e}rateur Lin{\'e}aire, IN2P3 CNRS et Universit{\'e} Paris-Sud, BP~34 91898 Orsay Cedex, France
\and Institut d'Astrophysique de Paris, INSU CNRS, 98~bis Boulevard Arago, 75014 Paris, France
\and Astronomical Observatory, Copenhagen University, Juliane Maries Vej 30, 2100 Copenhagen, Denmark
\and Coll{\`e}ge de France, PCC, IN2P3 CNRS, 11 place Marcelin Berthelot, 75231 Paris Cedex, France
\and Universidad de la Serena, Facultad de Ciencias, Departamento de Fisica, Casilla 554, La Serena, Chile
\and Department of Astronomy, Ohio State University, Columbus, OH 43210, U.S.A.
\and Physics Department, Ohio State University, Columbus, OH 43210, U.S.A.
\and Observatoire de Marseille, 2 place Le Verrier, 13248 Marseille Cedex 04, France
}
\offprints{Florian.Bauer@cea.fr}
\date{Received xx xx, 1999; accepted xx xx, 1999}
\maketitle
\markboth{The EROS collaboration: A catalog of Cepheids in the Magellanic Clouds}{}
\begin{abstract}
We present a catalog containing 
290 LMC and 590 SMC Cepheids which have been obtained
using the two 4k $\times$ 8k CCD cameras of 
the EROS~2 microlensing survey.
The Cepheids were selected from 1,134,000 and 504,000 stars in the central 
regions of the LMC and SMC respectively, that were monitored over 150 nights 
between October 1996 and February 1997, at a rate of one measurement
every night. For each Cepheid the light curves, period, magnitudes in the EROS~2 filter system,
Fourier coefficients, J2000 coordinates and cross-identifications 
with objects referenced in the CDS Simbad database
are presented. Finding charts of identified Cepheids in clusters NGC 1943, NGC 1958 and Bruck 56
are presented. The catalogue and the individual light--curves
will be electronically available through the CDS (Strasbourg). 
\end{abstract}
\section{Introduction}\label{intro}
The microlensing search projects towards the Magellanic Clouds have the ability to nightly 
monitor millions of stars over long time-scales, and are therefore 
suitable to detect and follow variable stars in a systematic 
way. The EROS~1 collaboration has already reported the discovery of 97 Cepheids 
towards the bar of the LMC (\cite{beaulieu1995}), and 450 Cepheids towards 
the SMC (\cite{beaulieu1996}). Owing to its larger spatial coverage the 
MACHO collaboration has reported the discovery of 1466 single mode and 73 
double-mode Cepheids towards the LMC (\cite{welch}). On the other hand the OGLE
collaboration reported the discovery of 1240 LMC and 2140 SMC Cepheids 
(\cite{udalski1}, \cite{udalski2}, \cite{udalski3}).

The present paper describes the results of a dedicated Cepheid campaign performed between 
October 1996 and February 1997 in order to allow a comparison between LMC and SMC 
Cepheids. The next section describes the observational setup.
In Sect.~3 the pipeline from the raw data to the Cepheid catalogue is 
presented with emphasis on the period search algorithm and F - 1-OT mode discrimination.
In Sect.~4 the EROS~2 filter system is characterized, while in Sect.~5 the astrometry
and the cross-identification with known objects will be described. Finally we present
the finding charts of three clusters (NGC 1943, NGC 1958, and Bruck 56)
containing Cepheids.
\section{Observations}\label{obs}
The observations were done using the EROS~2 experimental setup mounted at the
La Silla Observatory in Chile and which has been operational since June 1996.
The setup consists of a 1m F/5 Ritchey--Chr\'{e}tien telescope, with 
two 2 $\times$ 4 CCD mosaic cameras mounted in 
different focal planes. The light is split by means of a dichroic cube, allowing 
simultaneous imaging in two passband,
ranging from 420--720 nm ($V_{\rm EROS}$) and 650--920 nm ($R_{\rm EROS}$)
\footnote{We stress that these filters are different from those of the EROS~1 programme (see e.g.
\cite{sasselov_metal}, \cite{grison}).}. 
The mean wavelengths are respectively at 560 and 760 nm respectively.
The two cameras comprises 8 thick CCDs (Loral 2K $\times$ 2K, 3 edge buttable) 
each. 
The pixel size is 15 $\mu$m, which translates into 0.6$\arcsec$ 
on the sky, since the telescope is equipped with
an focal reducer yielding an aperture of F/5.
The total field of view covers 
an area of 0.7$\degr$ (right ascension) $\times$ 1.4$\degr$ (declination).
The acquisition system is able to treat up to 20 Gbytes of raw data every
night. It consists of two VME crates (one per color), which manage the different
real-time tasks, and a pool of two Alpha stations, where images are flat--fielded 
after a check of image quality. The raw and reduced data are finally saved on DLT tapes 
and shipped to France. The final data processing is done 
at the CCPN--IN2P3 national computing facility in Lyons.
A detailed description of the EROS~2 
experimental setup can be found in \cite{bauer1} or \cite{bauer2}.\\
\begin{table}
\caption{Coordinates (J2000) of the observed fields. The $N_{color}$ rows give 
the number of accumulated images for each field. }\label{tabel_champ}
\begin{tabular}{l|c|c|c|c}
\hline
field            &   LMC1     &  LMC2       &  SMC1      &  SMC2       \\
\hline
$\alpha$         & 05:23:12.8 & 05:15:14.8  & 00:51:39.1 & 00:41:39.2  \\
$\delta$         &-69:46:14.7 &-69:46:10.9  &-73:37:18.9 &-73:43:13.9  \\ 
$N_{blue}$       &    160     &   162       &   108      &  109        \\
$N_{red}$        &    162     &   167       &   117      &  111        \\
\hline
\end{tabular}
\end{table}
From October 1996 to February 1997, two fields per Magellanic Cloud 
(see Table 1) were monitored about once per night with an exposure time
of 21 seconds. The main goal of these observations 
being to get two comparable sets of LMC and SMC Cepheids, 
particular efforts has been paimade to observe the fields 
almost simultaneously and at same airmass (TS = 2h40).
First results from this campaign were reported in \cite{bauer3}.
\section{Data processing}\label{data}
\subsection{Photometry}\label{data_photo}
We used the standard EROS~2 photometric package PEIDA++, which
allows a rapid and efficient treatment of the large amount of data 
from the microlensing search program (\cite{ansari}).\\ 
The first step of the photometric data-reduction pipeline
consists of building a reference star catalog in both colors using a high quality
image (hereafter called reference image). The star detection is done
using our custom made package CORRFIND (\cite{pallanque}).
The final reference catalog contains 
the pixel position and the flux for each of the detected stars. \\
The data-reduction pipeline for the subsequent images 
(hereafter called current image) is the following:
\begin{enumerate}
\item Geometrical alignment of the current image relative
to the reference image, using the position of a few dozen
bright isolated stars. 
\item Determination of the PSF, which is specific to the current image, using 
bright isolated stars.  
\item Each star contained in the reference catalog is fitted
on the current image,
by fixing the position and taking into account the neighboring stars.
\item The average flux of the current image stars 
is photometrically realigned relative to the average flux in the reference
catalog, assuming that the majority of stars do not vary.
\item The photometric errors for each star of the current image
are calculated using the PSF fitting errors, combined
with the magnitude dispersion of stars with comparable flux. 
\end{enumerate}
Using the PEIDA++ photometry package 
we were able to reconstruct the light curves of
1,134,000 LMC and 504,000 SMC stars.\\
Two red CCDs (called \#R1 and \#R2) misfunctioned during the observation period.
Cepheid detection was still possible on the blue counterpart CCD, but no color
was available. 
As our main goal was to get a clean set of Cepheid data 
we concentrated our analysis on the remaining 
2$\times$6 CCDs.
\subsection{Search for variable stars}\label{data_var}
\begin{figure}
\hbox{\epsfig{file=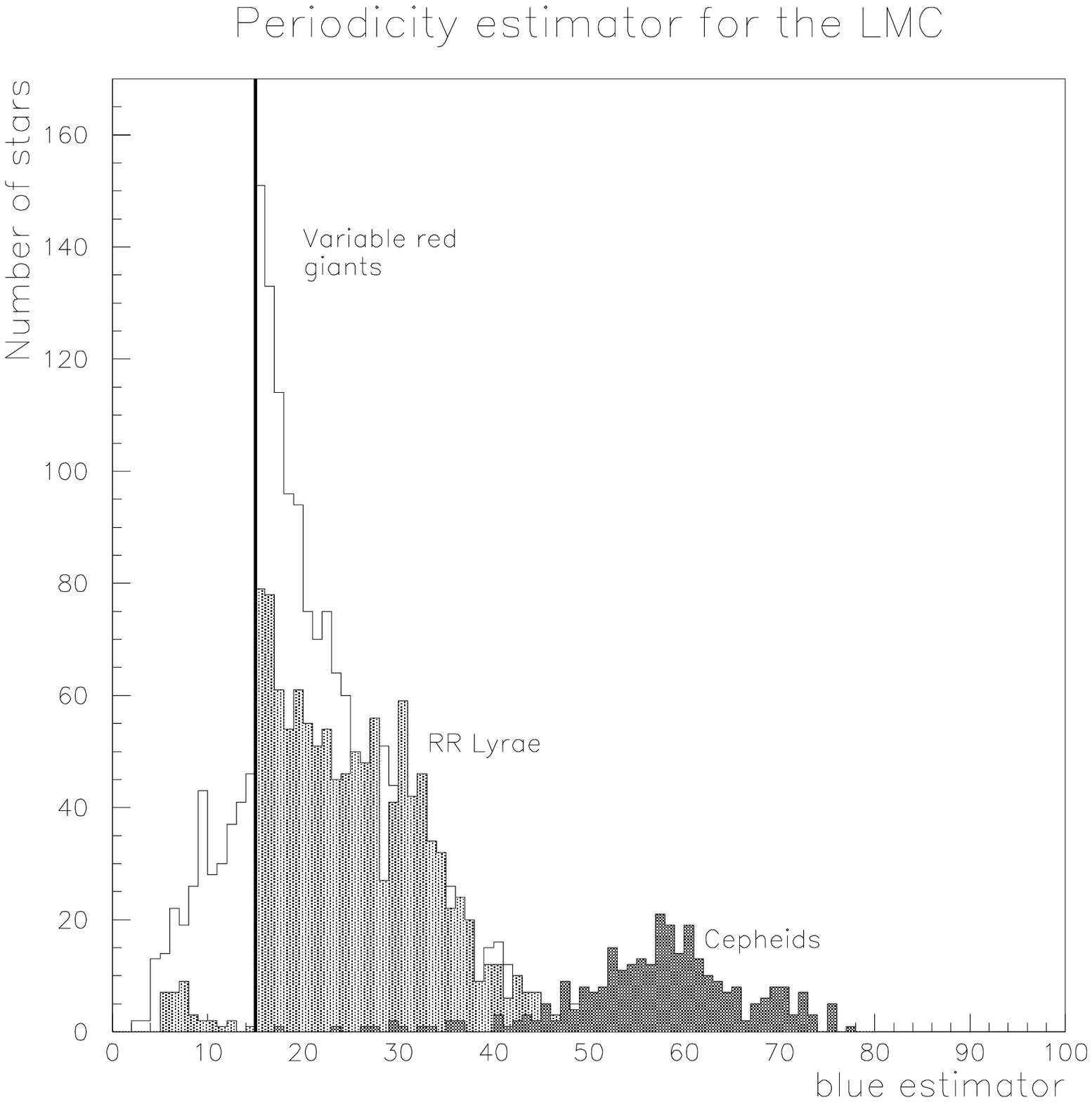,width=8.0cm}}  
\hbox{\epsfig{file=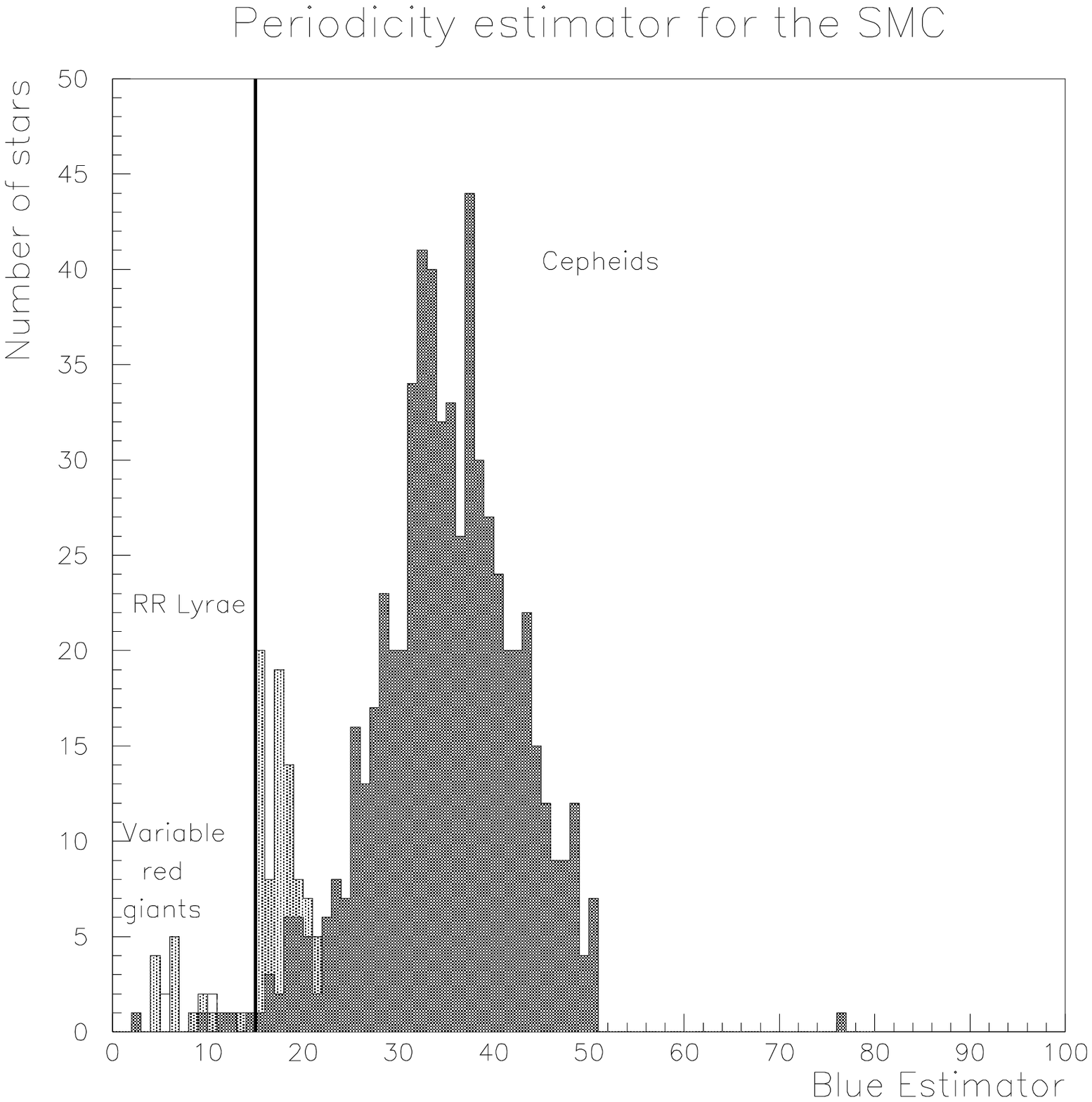,width=8.0cm}}
   \caption{Distribution of the periodicity search estimator for three different kinds of variable
stars. Higher values of the estimator are more likely to be variable stars.
The vertical line represents the cut we applied. The Cepheid distribution does not appear to 
be truncated and to has a good signal to noise ratio.}\label{estimator}
\end{figure}

Once the photometry was completed, we used an algorithm similar to \cite{scargle} 
to extract periodic light curves.  
This algorithm has the advantage of giving a periodicity estimator E, 
which can be easily converted to a probability of false detection,
allowing us to estimate the completeness of the data-set.
A total of 5497 LMC and 1929 SMC variable stars were detected.
In Fig.~\ref{variables} three diagrams are shown for each Cloud. 
At the top, the Color--Magnitude diagrams exhibit three main populations. 
The red giant variables are on the right, with colors ranging from 
$\sim$ 0.8 to 1.8 and  $V_{\rm EROS}\sim 15-17$ mag. 
The RR-Lyrae are located at colors ranging from 
$\sim$ 0.1 to $\sim$ 0.8 and $V_{\rm EROS}<18$~mag in the lMC
and $V_{\rm EROS}<18.5$ in the SMC,
while the Cepheids are in the color domain 
$\sim$ 0.2--0.7 with $V_{\rm EROS} \sim 14.5-18$ mag  in the LMC 
and $V_{\rm EROS} \sim 14-18.5$ mag  in the SMC. 
The Period--Luminosity diagrams (in the middle) 
of Fig.~\ref{variables} show two vertical clouds of points at $P=1$ day. They 
are likely to be due to long-period red giants for which aliases at $1+\nu$ or $1-\nu$
days have been detected. The RR-Lyrae are visible at the lower left 
corner. This indicates that the experiment is sensitive to short-period 
variations. Therefore the Cepheid sample is not truncated, as confirmed 
in Fig.~\ref{estimator} where the distribution of the periodicity-search 
estimators in the blue color are plotted for each of the three populations 
discussed above. A star was declared variable when its periodicity 
estimator was greater than 15. For variable stars detected in both colors, this 
corresponds to a negligeable statistical background of 10$^{-8}$.
On the other hand we estimate that the detection efficiency is 
about 80 $\%$, due to dead zones in the CCD.
\subsection{Cepheid selection}\label{data_cep}
The Cepheids were selected 
using wide cuts in the Period--Luminosity and Color--Magnitude planes, which 
for the LMC are:
\begin{eqnarray}
18.0 + 2.9 \log(P) < V_{\rm EROS} < 15.6 + 2.9 \log(P) \\
18.0 + 2.9 \log(P) < R_{\rm EROS} < 16.0 + 2.9 \log(P) \\
  V_{\rm EROS} < 17.6 \\
0.2 < V_{\rm EROS}-R_{\rm EROS} < 0.8
\end{eqnarray}
and in the SMC:
\begin{eqnarray}
18.3 + 2.9 \log(P) < V_{\rm EROS} < 16.0 + 2.9 \log(P) \\
18.9 + 2.9 \log(P) < R_{\rm EROS} < 16.4 + 2.9 \log(P) \\
  V_{\rm EROS} < 18.8 \\
0.2 < V_{\rm EROS}-R_{\rm EROS} < 0.8
\end{eqnarray}
In order to avoid pollution by eclipsing binaries and other objects 
all the Cepheid light curves were visually inspected.
The rejection fraction of this step was 40$\%$ in the LMC and 35$\%$
in the SMC.
\subsection{Fourier analysis of the light curves}\label{data_four}
In the past Fourier analysis of Cepheid light curves has prooved to be a powerful
tool for the comparison of hydrodynamic model predictions with observations.
Following (\cite{simonlee}), we adopt 
a Fourier analysis of the form:
\begin{equation}
m = m_{0} + \sum^{N}_{l=1} a_{l} cos(\frac{2 \pi}{P} l t + \phi_{l}),
\end{equation}
where $P$ is the period, $\phi$ the phase and $a_{l}$ the amplitude. 
The amplitude ratio
\begin{equation}
R_{kl} = \frac{a_{k}}{a_{l}}  (k>l)
\end{equation}
reflects the asymmetry of the light-curves with respect 
to a sinusoidal curve, which should typically have $R_{21}=0$, 
while the phase difference is:
\begin{equation}
\phi_{kl} = \phi_{k} - k \phi_{l}  \quad \quad \quad (k>l)
\end{equation}
%
%gives the full width at half maximum of the light--curve.
($\phi_{kl}$ is defined modulo 2$\pi$).\\
\begin{figure}
  \hbox{\epsfig{file=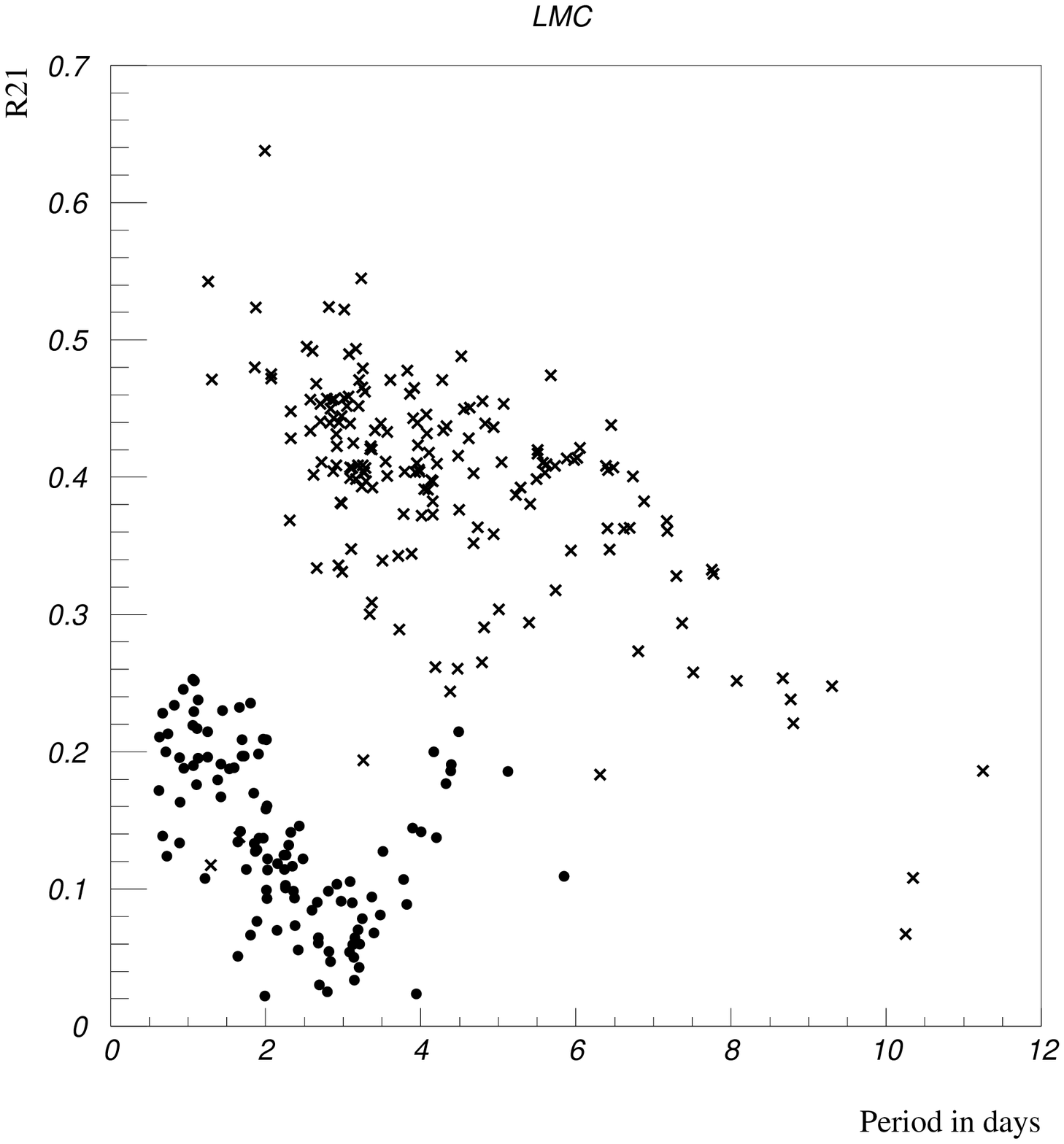,width=8.0cm}}  
  \hbox{\epsfig{file=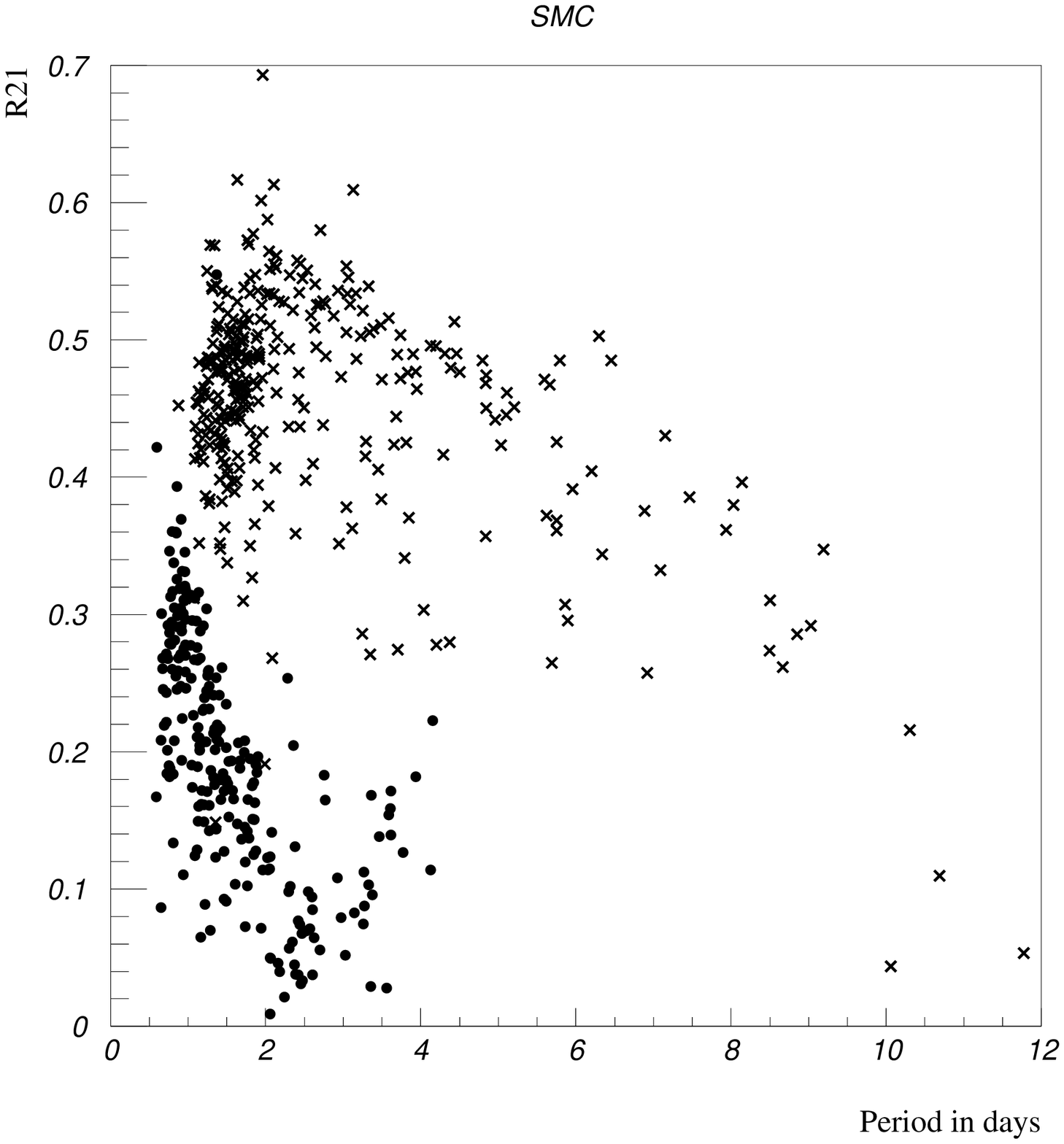,width=8.0cm}}
  \caption{Distinction between F and 1-OT~Cepheids in the $R_{21}$-Period plane. }\label{dist}
\end{figure}
The Fourier coefficients, the periods and the magnitude
of individual Cepheid light curves were determined using a fit with 5 harmonics.
The reported errors on the various Fourier cofficients were obtained from these fits.\\
A comparison of the Cepheid periods in the present catalogue
with the period found in the EROS~1 database for the same Cepheids suggests
a frequency uncertainty of $\sigma_{\nu} = 3.5 \ 10^{-4}$ days$^{-1}$.
The magnitude in the present catalog are mean--intensity magnitudes
($m = -2.5 \log(\frac{1}{N} \sum_{i}^{N} flux_{i})$).
The peak to peak amplitudes have been calculated using 
the Fourier coefficients resulting from the fit.
The various $R_{kl}$-period, $\phi_{kl}$-period and amplitude-period planes are 
plotted in Fig.\ref{r_1f}, Fig.\ref{r_1ot} and Fig.\ref{amplitude}.
The various period distributions are plotted in Fig.\ref{period}.
\subsection{Distinction between fundamental mode and first overtone Cepheids}\label{data_dist}
The distinction between F and 1-OT~Cepheids 
was done using the $R_{21}$ coefficient--Period plane.
In Fig.~\ref{dist} both populations are clearly visible: the Cepheids 
with a low $R_{21}$ are the 1-OT~Cepheids, those with a high $R_{21}$ are the F~Cepheids.

As a first aproximation we used the function $f(P)=0.4 + P/(25$ days 
in order to separe the two Cepheid populations in the $R_{21}$--Period 
plane, which resulted in a clear separation of two 
different sequences in the Period--Luminosity plane (see \cite{bauer3}). 
A second visual inspection of the Cepheid light curves was then achieved, in order
to classify the few stars which lie in between the F and 1-OT~Cepheid samples.

We found that 38 $\%$ of LMC and 41 $\%$ of SMC Cepheids were 1-OT~Cepheids, with 
periods ranging from 0.5 to 6 days. Table \ref{number} summarizes the different types 
of stars obtained from the bulk of the data.
\begin{table}[h]
\begin{center}
   \caption{Number of detected stars, of detected variables and of Cepheids}\label{number}
\begin{tabular}{l|cc|cc}
\hline
     &   Detected stars  & Variables  &  F          &   1-OT        \\        
\hline
LMC  &      1.134.000    &  5497      &   177        &   113        \\
SMC  &        504.000    &  1929      &   351        &   239        \\
\hline
\end{tabular}
\end{center}
\end{table}
\section{Photometric calibration}\label{calib}
The main scientific goal of EROS~2 is the search for microlensing events.
This kind of survey does not requiire absolute photometry 
performed with standard filter systems.
Therefore the two EROS~2 filter passbands have been chosen 
to be wide in order to shorten the exposure times by nearly a factor of 2 with 
respect to more standard filters. 
This means that twice many as many fields can be followed every night and the 
probability to detect ongoing microlensing events doubles.
On the other hand the drawback of this choice is that the 
mean Cepheid magnitudes in the present catalogue are given in the
EROS~2 non--standard filter system. \\

In order to connect the EROS~2 filter system to the Johnson-Cousins one,
we determined tertiary standards in BVRI, using 
the Danish 1.4m telescope at La Silla. Two data-sets have been analyzed, 
one centered on CCD $\#$5 of field LMC~1 ($\sim$ 8000 stars) and the 
other centered on CCD $\#$6 of field LMC~1 ($\sim$ 5000 stars). 

Using these two data sets, linear fit of the various color-color 
planes gave the following conversion equations:
\begin{eqnarray}\label{coloreq}
V_{j} = V_{\rm EROS} + 0.33(3) \ (V_{\rm EROS}-R_{\rm EROS}) \\
R_{c} = V_{\rm EROS} - 0.38(3) \ (V_{\rm EROS}-R_{\rm EROS}) \\
I_{c} = V_{\rm EROS} - 1.03(3) \ (V_{\rm EROS}-R_{\rm EROS})  
\end{eqnarray}
where the color term uncertainty corresponds to the difference
between the color terms of both data sets. 
These uncertainties are close to the errors given by the fit.
The residuals of these fits have a typical FWHM of 0.4~mag.\\
The EROS~2 magnitudes $V_{\rm EROS}$ and $R_{\rm EROS}$ were measured to be:
\begin{eqnarray}
V_{\rm EROS} = -2.5 \log( \frac{\phi_{V}}{t} ) + 22.46(6) + I_{i} \\
R_{\rm EROS} = -2.5 \log( \frac{\phi_{R}}{t} ) + 21.65(7) + I_{i}
\end{eqnarray}\label{pzero}
$\phi$ beeing the measured flux in ADU and $t$ the exposure time (21~seconds).
The errors on the offset have been determined using the dispersion of
the results obtained in the various color-color planes of both data sets.\\
The offsets between the different CCDs have been taken into account,
applying inter-CCD corrections $I_{i}$ to our Cepheids (see Table 3). 
This inter-CCD corrections have been determined by taking a field 
with each of the CCD pairs and applying the same photometric reduction chain. 
We estimate the accuracy of these inter-CCD corrections to be 0.03~mag. 
\begin{table}[h]
\begin{center}
   \caption{The applied inter--CCD corrections I$_{i}$.}\label{number}
\begin{tabular}{lllllllll}
\hline
         &  0     &  1    &  2   &  3   &  4   &  5    &  6   &  7\\    
\hline
%V$_{e}$  &  22.36 & 22.20 & 22.49 & 22.46 &  22.48 & 22.43 & 22.39 & 22.37 \\
%R$_{e}$  &  21.63 & 21.67 & 21.72 & 21.65 &  21.63 & 21.67 & 21.61 & 21.56 \\
V$_{e}$ & -0.10 & -0.26 & 0.03 & 0.00 &  0.02 & -0.03 & -0.07 & -0.09\\
R$_{e}$ & -0.02 &  0.02 & 0.07 & 0.00 & -0.02 &  0.02 & -0.04 & -0.09\\  
\hline
\end{tabular}
\end{center}
\end{table}

Finally we stress that, if the reader intends to apply Eq.\ref{coloreq}
to the offset of the fitted PL--relations reported in Table 2 of \cite{bauer3}, 
a systematic correction of $\Delta V_{\rm EROS}=$-0.35~mag 
and $\Delta R_{\rm EROS}=$-0.56~mag has to be added to the offset. 
\section{Astrometry, cross-identification and clusters}\label{astro}
The J2000 equatorial coordinates (J2000 equinox) of individual stars have been 
obtained as follows. First, we computed the suitable WCS keywords 
into the header of the EROS~2 reference images
using the WCSTools package from Center for Astrophysics 
(Harvard, USA). Then the images were mapped with the MACS 
catalog (\cite{tucholke}), which presents several advantages:
{\it (i)} it is dedicated to Magellanic Clouds; {\it (ii)} it refers to FK5 system; {\it (iii)} 
it contains only stars undisturbed by close neighbors; {\it (iv)} 
distorsions of ESO Schmidt plates have been taken into account, as an end
result internal errors remain well below the size of an EROS~2 pixel.

When possible, the cross-identification of each star with previously 
kwown identifiers has been done using the batch utilities of the Simbad 
database available at the CDS. The search for objects was performed 
within a radius of 1 arcmin and the star selection within 10 arcsec.
We found 26 Cepheids which could be cross--identified with the published EROS~1
Cepheid catalogue (\cite{beaulieu1995}).  

The identifier of each star is built following the recommendations of 
the IAU Commission 5 in The Rules and Regulations for 
Nomenclature (see the Annual Index of A\&A). The general acronym of the 
catalog is EROS2C followed by J2000 equatorial coordinates 
in the format J$HHMM\pm DDMM$. The next item is the four-digit numeral 
position of the star in the catalog, where the leftmost digit gives the  
Cepheid type: (3) LMC F~Cepheid; (4) LMC 1-OT~Cepheid, (5) SMC F~Cepheid; (6) SMC 1-OT~Cepheid. 
The remainder of the identifier in parentheses gives some information relating to the 
internal organisation of the EROS database : Cp$n$ is the name of the 
field (1--2 for LMC, 3--4 for SMC), followed by the number of the CCD and 
the location on the image (divided into 4 quarters k, l, m, and n). The 
last number is the star identifier used in the EROS database.
\begin{table}
\caption{Parameters of the Cepheids found in NGC 1943, NGC 1958 and Bruck 56. }\label{cluster}
\begin{tabular}{rcccl}
\hline
            &{\bf NGC}         & {\bf 1943}     &                     &                           \\
\hline
N$^{\circ}$ & $V_{\rm EROS}$   & $R_{\rm EROS}$ &  P (in days)        & remarks                   \\ 
\hline 
\hline 
4055        &  15.909	       &  15.597        &  2.0145             & 1-OT~Cepheid              \\
3024        &  15.550          &  15.370        &  2.8140             &                           \\
3036        &  16.000	       &  15.666        &  2.9691             &                           \\
3052        &  15.665	       &  15.383        &  3.1672             &                           \\
3056        &  15.376	       &  14.982	&  3.2324             & blended by 3114           \\
3064        &  15.719	       &  15.370	&  3.2800             &                           \\
3085        &  15.808	       &  15.439        &  3.8593             & blended                   \\
3114        &  15.523	       &  15.003	&  4.5257             & blended by 3056           \\
\hline
\hline
            &{\bf NGC }        & {\bf 1958}     &                     &                           \\
\hline
N$^{\circ}$ & $V_{\rm EROS}$   & $R_{\rm EROS}$ &  P (in days)        & remarks                   \\  
\hline 
\hline 
3002        &  17.322	       &  16.699        &  1.2950             & intercept?                \\
4085        &  15.137	       &  14.741	&  2.9766	      & 1-OT~Cepheid, intercept?  \\
3130        &  15.183	       &  14.702        &  5.2248             &                           \\
3132        &  15.120	       &  14.505        &  5.3957             &                           \\
3144        &  14.992	       &  14.471        &  5.9353             &                           \\
3145        &  15.371	       &  14.800	&  5.9763             &                           \\
3150        &  15.081	       &  14.523        &  6.4050             &                           \\
3155        &  15.183	       &  14.595	&  6.6204             &                           \\
3170        &   14.997	       &  14.395        &  8.8008             & visible resonance         \\
\hline 
\hline 
            &{\bf Bruck}       & {\bf 56}       &                     &                           \\
\hline
N$^{\circ}$ & $V_{\rm EROS}$   & $R_{\rm EROS}$ &  P (in days)  & remarks                         \\  
\hline 
\hline 
5009        &  17.468	       &  17.106	 &   1.1339           &                           \\
6139        &  16.719	       &  16.375	 &   1.4395           &   1-OT~Cepheid            \\
5102        &  17.216	       &  16.922	 &   1.5799           &                           \\
5178        &  17.569	       &  17.131	 &   1.9127           &                           \\
5338        &  14.258	       &  13.793	 &  15.6852           &                           \\
\hline 
\end{tabular}
\end{table}

In the present catalog we detected
three Cepheid overdensities centered around young clusters.
Such systems have already been observed in the past (\cite{efremov},
\cite{tsvetkov} and references therein) and are particularly interesting,
to test stellar evolution models or to measure the age of the cluster.
The parameters of these Cepheids are indicated in Table \ref{cluster}.
The identification maps are shown in Fig. \ref{cluster1}, \ref{cluster2}
and \ref{cluster3}.

The first cluster is known as NGC 1943 and 
seven F~Cepheids and one 1-OT~Cepheid have been found
in this direction. The F~Cepheids
have periods ranging between 2.8 and 4.5 days. The 1-OT~Cepheid has a period
of 2.01 days, which corresponds to the period of 2.8 days for a F~Cepheid
using the 1-OT and F~Cepheid period ratio of 0.7 (\cite{welch}).\\
The second cluster known as NGC 1958 contains 9 Cepheids.
Six of them are F~Cepheids with periods between
5.2 and 6.6 days. One F~Cepheid ($\#$3170) has a period of 8.8 days. 
The last two ($\#$3002 and $\#$4085) have a period
of 1.25 and 2.97 days (corespondig to a 4.16 days F~Cepheid). 
As thess periods are small compared to the others,
it is unclear if this Cepheid belongs to the cluster or not.
As the covered field has the dimensions 1.5 arcminutes$ \times$ 1.5~arcminutes it can be shown 
using the spatial density of all our LMC Cepheids 
that aproximatively 1 Cepheid is expected to lie randomly in this direction.\\
A third Cepheid cluster known as Bruck 56 has been found towards the SMC, with
four F~Cepheids and one 1-OT~Cepheid. One of the F~Cepheids ($\#$5343) has a period
of 15.68 days and has been detected in the core of the cluster. Thus it seems
improbable that this Cepheid is just an interloper. 
\section{Conclusion }
Using the EROS~2 microlensing detector we compiled a SMC and LMC Cepheid catalog
containing the light curves, Fourier coefficients, magnitudes, periods and
J2000 positions of 290 LMC and 590 SMC Cepheids.  
\begin{acknowledgements}
The authors are particulary grateful to the ESO staff at the La Silla Observatory
for their night and day assistance, and to the Observatoire de Haute Provence technical staff 
for their help in refurbishing and mounting 
the Marly telescope. We are grateful to the DAPNIA technical
staff for the maintenance of the CCD cameras.
We wish to thank J.F. Lecointe for his useful assistance during the mounting of online
acquisition system at La Silla, and the CCPN--IN2P3 staff in Lyons for their help during this first
EROS~2 data mass production. The WCSTools package was made available to us thanks
to the work of Doug Mink.
The Skycat/Gaia tool results mainly from joint efforts by Allan Brighton at 
ESO Garching, and Peter Draper of the Starlink Project, UK.
\end{acknowledgements}
{\sc \clearpage
 \begin{table}[pth] 
\tiny 
\begin{flushleft} 
\begin{tabular}{|l@{  }l@{  }l@{  }l@{ }l@{ }r@{ }r@{ }r@{ }r@{ }l@{ }l@{ }l@{ }l@{ }l@{ }l@{ }l@{ }l@{ }l|} 
\hline 
 & & & & & & & & & & & & & & & & & Xid, Cluster,  \\ 
EROS2C N$^{\circ}$ & Identification      & $\alpha$ \quad & $\delta$  \quad \quad \quad  \quad &  Period      &  $\langle V_{E} \rangle $  &   $\Delta V_{E}$            &               $\langle R_{E} \rangle $  &    $\Delta R_{E}$          &  $R_{21}$   &  $\sigma_{R_{21}}$   &        $R_{31}$ &   $\sigma_{R_{31}}$  &       $\phi_{21}$  &  $\sigma_{\phi_{21}}$        &  $\phi_{31}$   &  $\sigma_{\phi_{31}}$      &   Remarks     \\ 
\hline
\hline
J0519--7011\#3001 & (Cp16k--10969) & 5:19:16.4 & -70:11:59.1 &   1.2578 &  17.25 &  0.79 &  16.80 &  0.50 &   0.542 &  0.033 &  0.383 &  0.032 &  4.329 &  0.081 &  2.335 &  0.117 &      \\ 
J0525--6949\#3002 & (Cp15m--11646) & 5:25:26.3 & -69:49:49.0 &   1.2950 &  17.32 &  0.28 &  16.70 &  0.20 &   0.117 &  0.036 &  0.050 &  0.037 &  4.000 &  0.329 &  2.465 &  0.742 & NGC 1958 1OT-cepheid ?  \\ 
J0517--6928\#3003 & (Cp23m--11639) & 5:17:49.9 & -69:28:29.0 &   1.3088 &  16.71 &  0.58 &  16.28 &  0.38 &   0.471 &  0.020 &  0.236 &  0.018 &  3.781 &  0.053 &  1.616 &  0.095 & EROS 2001    \\ 
J0521--6958\#3004 & (Cp14n--10800) & 5:21:32.0 & -69:58:49.3 &   1.6636 &  16.80 &  0.23 &  16.16 &  0.15 &   0.138 &  0.044 &  0.054 &  0.044 &  4.304 &  0.343 &  3.756 &  0.839 & 1OT-cepheid ?    \\ 
J0514--6919\#3005 & (Cp20n--15692) & 5:14:01.3 & -69:19:51.2 &   1.8580 &  16.61 &  0.91 &  16.25 &  0.66 &   0.480 &  0.015 &  0.343 &  0.014 &  3.990 &  0.038 &  1.887 &  0.055 &      \\ 
J0525--6927\#3006 & (Cp13m--7388) & 5:25:19.3 & -69:27:05.3 &   1.8742 &  16.66 &  0.87 &  16.21 &  0.61 &   0.524 &  0.017 &  0.362 &  0.017 &  3.973 &  0.044 &  1.933 &  0.063 &      \\
\hline    
\end{tabular}
\end{flushleft} 
\end{table} 
\clearpage 
\clearpage 
 \begin{table}[pth] 
\tiny 
\begin{flushleft} 
\begin{tabular}{|l@{  }l@{  }l@{  }l@{ }l@{ }r@{ }r@{ }r@{ }r@{ }l@{ }l@{ }l@{ }l@{ }l@{ }l@{ }l@{ }l@{ }l|} 
\hline 
 & & & & & & & & & & & & & & & & & Xid, Cluster,  \\ 
EROS2C N$^{\circ}$ & Identification      & $\alpha$ \quad & $\delta$  \quad \quad \quad  \quad &  Period      &  $\langle V_{E} \rangle $  &   $\Delta V_{E}$            &               $\langle R_{E} \rangle $  &    $\Delta R_{E}$          &  $R_{21}$   &  $\sigma_{R_{21}}$   &        $R_{31}$ &   $\sigma_{R_{31}}$  &       $\phi_{21}$  &  $\sigma_{\phi_{21}}$        &  $\phi_{31}$   &  $\sigma_{\phi_{31}}$      &   Remarks     \\ 
\hline
\hline
J0526--6950\#3101 & (Cp15m--13683) & 5:26:31.7 & -69:50:25.9 &   4.1331 &  15.38 &  0.66 &  14.96 &  0.47 &   0.398 &  0.014 &  0.192 &  0.013 &  4.371 &  0.038 &  2.596 &  0.075 & EROS 2055    \\ 
J0522--7009\#3102 & (Cp16m--8817) & 5:22:44.1 & -70:09:39.5 &   4.1524 &  15.70 &  0.55 &  15.25 &  0.39 &   0.397 &  0.024 &  0.196 &  0.023 &  4.538 &  0.070 &  2.912 &  0.131 & NGC 1943    \\ 
J0526--6944\#3103 & (Cp13n--28576) & 5:26:28.9 & -69:44:17.2 &   4.1557 &  15.38 &  0.63 &  14.97 &  0.43 &   0.383 &  0.020 &  0.161 &  0.018 &  4.345 &  0.061 &  2.386 &  0.132 & HV 962    \\ 
J0526--7015\#3104 & (Cp17m--25045) & 5:26:59.5 & -70:15:46.9 &   4.1571 &  15.55 &  0.57 &  15.06 &  0.41 &   0.373 &  0.019 &  0.170 &  0.019 &  4.431 &  0.065 &  2.516 &  0.123 &      \\ 
J0521--6920\#3105 & (Cp10n--18908) & 5:21:46.7 & -69:20:41.2 &   4.1907 &  15.27 &  0.31 &  14.79 &  0.22 &   0.261 &  0.036 &  0.067 &  0.034 &  3.856 &  0.140 &  1.305 &  0.503 &      \\ 
J0518--6939\#3106 & (Cp23n--15648) & 5:18:50.9 & -69:39:12.2 &   4.2055 &  15.38 &  0.68 &  15.06 &  0.48 &   0.410 &  0.015 &  0.184 &  0.013 &  4.330 &  0.040 &  2.459 &  0.080 & HV 12010    \\
\hline    
\end{tabular}
\end{flushleft} 
\end{table} 
\clearpage 
\clearpage 
 \begin{table}[pth] 
\tiny 
\begin{flushleft} 
\begin{tabular}{|l@{  }l@{  }l@{  }l@{ }l@{ }r@{ }r@{ }r@{ }r@{ }l@{ }l@{ }l@{ }l@{ }l@{ }l@{ }l@{ }l@{ }l|} 
\hline 
 & & & & & & & & & & & & & & & & & Xid, Cluster,  \\ 
EROS2C N$^{\circ}$ & Identification      & $\alpha$ \quad & $\delta$  \quad \quad \quad  \quad &  Period      &  $\langle V_{E} \rangle $  &   $\Delta V_{E}$            &               $\langle R_{E} \rangle $  &    $\Delta R_{E}$          &  $R_{21}$   &  $\sigma_{R_{21}}$   &        $R_{31}$ &   $\sigma_{R_{31}}$  &       $\phi_{21}$  &  $\sigma_{\phi_{21}}$        &  $\phi_{31}$   &  $\sigma_{\phi_{31}}$      &   Remarks     \\ 
\hline
\hline
J0512--6957\#4001 & (Cp24l--4327) & 5:12:22.8 & -69:57:31.7 &   0.6259 &  17.56 &  0.33 &  17.14 &  0.21 &   0.172 &  0.037 &  0.212 &  0.038 &  3.091 &  0.225 &  0.737 &  0.203 &      \\ 
J0520--6909\#4002 & (Cp10k--14852) & 5:20:54.3 & -69:09:56.9 &   0.6286 &  17.21 &  0.25 &  16.97 &  0.21 &   0.211 &  0.091 &  0.070 &  0.090 &  2.187 &  0.462 &  0.757 &  2.058 &      \\ 
J0527--6947\#4003 & (Cp15m--6547) & 5:27:00.7 & -69:47:45.4 &   0.6740 &  17.17 &  0.20 &  16.89 &  0.21 &   0.228 &  0.052 &  0.109 &  0.053 &  3.558 &  0.272 &  2.038 &  0.530 &      \\ 
J0522--6954\#4004 & (Cp14m--22725) & 5:22:14.1 & -69:54:38.1 &   0.6749 &  17.23 &  0.19 &  16.63 &  0.13 &   0.139 &  0.070 &  0.185 &  0.068 &  3.096 &  0.476 &  0.712 &  0.435 &      \\ 
J0522--7021\#4005 & (Cp16n--10908) & 5:22:26.0 & -70:21:45.3 &   0.7112 &  17.34 &  0.29 &  17.04 &  0.27 &   0.200 &  0.055 &  0.147 &  0.053 &  3.287 &  0.267 &  0.977 &  0.383 &      \\ 
J0526--6952\#4006 & (Cp15m--18259) & 5:26:02.0 & -69:52:09.4 &   0.7292 &  17.01 &  0.24 &  16.62 &  0.16 &   0.124 &  0.044 &  0.055 &  0.043 &  3.774 &  0.335 &  6.181 &  0.802 &      \\
\hline    
\end{tabular}
\end{flushleft} 
\end{table} 
\clearpage 
\clearpage 
 \begin{table}[pth] 
\tiny 
\begin{flushleft} 
\begin{tabular}{|l@{  }l@{  }l@{  }l@{ }l@{ }r@{ }r@{ }r@{ }r@{ }l@{ }l@{ }l@{ }l@{ }l@{ }l@{ }l@{ }l@{ }l|} 
\hline 
 & & & & & & & & & & & & & & & & & Xid, Cluster,  \\ 
EROS2C N$^{\circ}$ & Identification      & $\alpha$ \quad & $\delta$  \quad \quad \quad  \quad &  Period      &  $\langle V_{E} \rangle $  &   $\Delta V_{E}$            &               $\langle R_{E} \rangle $  &    $\Delta R_{E}$          &  $R_{21}$   &  $\sigma_{R_{21}}$   &        $R_{31}$ &   $\sigma_{R_{31}}$  &       $\phi_{21}$  &  $\sigma_{\phi_{21}}$        &  $\phi_{31}$   &  $\sigma_{\phi_{31}}$      &   Remarks     \\ 
\hline
\hline
J0521--7008\#4101 & (Cp16m--6827) & 5:21:26.6 & -70:08:39.0 &   3.7762 &  15.19 &  0.26 &  14.76 &  0.18 &   0.107 &  0.042 &  0.033 &  0.041 &  3.336 &  0.374 &  1.112 &  1.202 &      \\ 
J0514--6921\#4102 & (Cp20n--18900) & 5:14:10.8 & -69:21:26.8 &   3.8199 &  14.75 &  0.10 &  14.47 &  0.09 &   0.089 &  0.070 &  0.029 &  0.074 &  2.927 &  0.877 &  2.510 &  2.516 &      \\ 
J0526--6927\#4103 & (Cp13m--7526) & 5:26:23.5 & -69:27:01.2 &   3.8948 &  15.01 &  0.30 &  14.62 &  0.20 &   0.144 &  0.032 &  0.057 &  0.034 &  3.416 &  0.232 &  0.970 &  0.587 &      \\ 
J0524--6950\#4104 & (Cp15k--13260) & 5:24:19.3 & -69:50:40.4 &   3.9404 &  14.69 &  0.13 &  14.19 &  0.08 &   0.023 &  0.069 &  0.073 &  0.065 &  2.213 &  2.631 &  1.330 &  0.908 &      \\ 
J0513--6921\#4105 & (Cp20n--19242) & 5:13:52.5 & -69:21:37.4 &   4.0022 &  14.70 &  0.29 &  14.38 &  0.20 &   0.142 &  0.043 &  0.046 &  0.061 &  3.175 &  0.322 &  0.919 &  1.374 &      \\ 
J0526--6951\#4106 & (Cp15m--16837) & 5:26:48.7 & -69:51:32.5 &   4.1673 &  14.70 &  0.31 &  14.37 &  0.21 &   0.200 &  0.027 &  0.076 &  0.024 &  3.670 &  0.126 &  0.630 &  0.346 & EROS 2057 NGC 1969 NGC 1971\\
\hline    
\end{tabular}
\end{flushleft} 
\end{table} 
\clearpage 
\clearpage 
 \begin{table}[pth] 
\tiny 
\begin{flushleft} 
\begin{tabular}{|l@{  }l@{  }l@{  }l@{ }l@{ }r@{ }r@{ }r@{ }r@{ }l@{ }l@{ }l@{ }l@{ }l@{ }l@{ }l@{ }l@{ }l|} 
\hline 
 & & & & & & & & & & & & & & & & & Xid, Cluster,  \\ 
EROS2C N$^{\circ}$ & Identification      & $\alpha$ \quad & $\delta$  \quad \quad \quad  \quad &  Period      &  $\langle V_{E} \rangle $  &   $\Delta V_{E}$            &               $\langle R_{E} \rangle $  &    $\Delta R_{E}$          &  $R_{21}$   &  $\sigma_{R_{21}}$   &        $R_{31}$ &   $\sigma_{R_{31}}$  &       $\phi_{21}$  &  $\sigma_{\phi_{21}}$        &  $\phi_{31}$   &  $\sigma_{\phi_{31}}$      &   Remarks     \\ 
\hline
\hline
J0053--7327\#5001 & (Cp33l--4323) & 0:53:18.9 & -73:27:32.0 &   0.8752 &  18.44 &  0.75 &  17.94 &  0.48 &   0.452 &  0.047 &  0.341 &  0.046 &  4.050 &  0.129 &  2.255 &  0.180 &      \\ 
J0052--7348\#5002 & (Cp35k--4271) & 0:52:09.2 & -73:48:19.4 &   1.0749 &  17.75 &  0.90 &  17.28 &  0.58 &   0.312 &  0.023 &  0.220 &  0.022 &  3.413 &  0.080 &  1.087 &  0.117 &      \\ 
J0052--7348\#5002 & (Cp35k--4271) & 0:52: 9.2 & -73:48:19.4 &   1.0749 &  17.75 &  0.90 &  17.28 &  0.58 &   0.312 &  0.023 &  0.220 &  0.022 &  3.413 &  0.080 &  1.087 &  0.117 &      \\ 
J0049--7307\#5003 & (Cp30n--8498) & 0:49:19.9 & -73:07:35.4 &   1.0884 &  17.81 &  0.70 &  17.42 &  0.45 &   0.413 &  0.036 &  0.260 &  0.035 &  3.802 &  0.107 &  1.319 &  0.164 &      \\ 
J0040--7403\#5004 & (Cp44n--4143) & 0:40:38.4 & -74:03:07.8 &   1.0970 &  17.99 &  0.85 &  17.66 &  0.66 &   0.437 &  0.027 &  0.249 &  0.024 &  3.781 &  0.073 &  1.582 &  0.124 &      \\ 
J0048--7313\#5005 & (Cp30l--18479) & 0:48:11.7 & -73:13:20.0 &   1.1117 &  17.77 &  0.45 &  17.22 &  0.29 &   0.455 &  0.100 &  0.276 &  0.090 &  3.741 &  0.267 &  1.108 &  0.427 &      \\
\hline    
\end{tabular}
\end{flushleft} 
\end{table} 
\clearpage 
\clearpage 
 \begin{table}[pth] 
\tiny 
\begin{flushleft} 
\begin{tabular}{|l@{  }l@{  }l@{  }l@{ }l@{ }r@{ }r@{ }r@{ }r@{ }l@{ }l@{ }l@{ }l@{ }l@{ }l@{ }l@{ }l@{ }l|} 
\hline 
 & & & & & & & & & & & & & & & & & Xid, Cluster,  \\ 
EROS2C N$^{\circ}$ & Identification      & $\alpha$ \quad & $\delta$  \quad \quad \quad  \quad &  Period      &  $\langle V_{E} \rangle $  &   $\Delta V_{E}$            &               $\langle R_{E} \rangle $  &    $\Delta R_{E}$          &  $R_{21}$   &  $\sigma_{R_{21}}$   &        $R_{31}$ &   $\sigma_{R_{31}}$  &       $\phi_{21}$  &  $\sigma_{\phi_{21}}$        &  $\phi_{31}$   &  $\sigma_{\phi_{31}}$      &   Remarks     \\ 
\hline
\hline
J0051--7301\#5101 & (Cp30m--14692) & 0:51:07.9 & -73:01:48.6 &   1.5742 &  16.95 &  0.97 &  16.72 &  0.73 &   0.485 &  0.024 &  0.348 &  0.023 &  3.938 &  0.062 &  1.769 &  0.090 & 1OT-cepheid ?    \\ 
J0050--7312\#5102 & (Cp30n--17359) & 0:50:56.6 & -73:12:25.8 &   1.5799 &  17.22 &  1.00 &  16.92 &  0.77 &   0.515 &  0.022 &  0.338 &  0.021 &  3.948 &  0.052 &  1.809 &  0.078 & Bruck 56    \\ 
J0046--7355\#5103 & (Cp34l--7143) & 0:46:48.1 & -73:55:13.1 &   1.5801 &  17.35 &  1.12 &  17.07 &  0.82 &   0.492 &  0.024 &  0.284 &  0.023 &  3.890 &  0.061 &  1.651 &  0.094 &      \\ 
J0055--7351\#5104 & (Cp35n--4662) & 0:55:55.9 & -73:51:33.3 &   1.5842 &  17.35 &  0.87 &  17.02 &  0.66 &   0.463 &  0.025 &  0.390 &  0.026 &  3.906 &  0.072 &  1.927 &  0.096 &      \\ 
J0045--7329\#5105 & (Cp43m--12038) & 0:45:14.4 & -73:29:26.7 &   1.5925 &  17.41 &  0.94 &  17.00 &  0.72 &   0.447 &  0.019 &  0.344 &  0.019 &  3.907 &  0.053 &  1.715 &  0.073 & Bruck 39    \\ 
J0049--7305\#5106 & (Cp30m--12958) & 0:49:14.9 & -73:05:28.1 &   1.5931 &  17.08 &  0.81 &  16.68 &  0.61 &   0.477 &  0.028 &  0.341 &  0.027 &  3.843 &  0.073 &  1.604 &  0.106 &      \\
\hline    
\end{tabular}
\end{flushleft} 
\end{table} 
\clearpage 
\clearpage 
 \begin{table}[pth] 
\tiny 
\begin{flushleft} 
\begin{tabular}{|l@{  }l@{  }l@{  }l@{ }l@{ }r@{ }r@{ }r@{ }r@{ }l@{ }l@{ }l@{ }l@{ }l@{ }l@{ }l@{ }l@{ }l|} 
\hline 
 & & & & & & & & & & & & & & & & & Xid, Cluster,  \\ 
EROS2C N$^{\circ}$ & Identification      & $\alpha$ \quad & $\delta$  \quad \quad \quad  \quad &  Period      &  $\langle V_{E} \rangle $  &   $\Delta V_{E}$            &               $\langle R_{E} \rangle $  &    $\Delta R_{E}$          &  $R_{21}$   &  $\sigma_{R_{21}}$   &        $R_{31}$ &   $\sigma_{R_{31}}$  &       $\phi_{21}$  &  $\sigma_{\phi_{21}}$        &  $\phi_{31}$   &  $\sigma_{\phi_{31}}$      &   Remarks     \\ 
\hline
\hline
J0046--7327\#5201 & (Cp43m--9322) & 0:46:04.2 & -73:27:13.6 &   2.1253 &  17.00 &  0.42 &  16.59 &  0.30 &   0.407 &  0.034 &  0.211 &  0.034 &  4.079 &  0.099 &  2.301 &  0.168 &      \\ 
J0038--7348\#5202 & (Cp44m--5027) & 0:38:58.6 & -73:48:32.1 &   2.1310 &  17.18 &  1.24 &  16.76 &  0.90 &   0.552 &  0.015 &  0.384 &  0.014 &  4.182 &  0.036 &  2.159 &  0.053 &      \\ 
J0049--7257\#5203 & (Cp30m--6955) & 0:49:11.1 & -72:57:09.6 &   2.1371 &  16.85 &  0.75 &  16.53 &  0.53 &   0.562 &  0.030 &  0.370 &  0.028 &  4.146 &  0.071 &  2.212 &  0.108 &      \\ 
J0049--7257\#5203 & (Cp30m--6955) & 0:49:11.1 & -72:57: 9.6 &   2.1371 &  16.85 &  0.75 &  16.53 &  0.53 &   0.562 &  0.030 &  0.370 &  0.028 &  4.146 &  0.071 &  2.212 &  0.108 &      \\ 
J0047--7312\#5204 & (Cp30l--17688) & 0:47:14.4 & -73:12:51.1 &   2.1387 &  16.98 &  0.50 &  16.62 &  0.40 &   0.461 &  0.045 &  0.206 &  0.043 &  4.131 &  0.127 &  2.341 &  0.239 & [H86] 93    \\ 
J0045--7334\#5205 & (Cp43n--4860) & 0:45:40.2 & -73:34:16.3 &   2.1538 &  16.69 &  1.08 &  16.40 &  0.80 &   0.502 &  0.013 &  0.294 &  0.013 &  3.986 &  0.037 &  1.868 &  0.058 &      \\
\hline    
\end{tabular}
\end{flushleft} 
\end{table} 
\clearpage 
\clearpage 
 \begin{table}[pth] 
\tiny 
\begin{flushleft} 
\begin{tabular}{|l@{  }l@{  }l@{  }l@{ }l@{ }r@{ }r@{ }r@{ }r@{ }l@{ }l@{ }l@{ }l@{ }l@{ }l@{ }l@{ }l@{ }l|} 
\hline 
 & & & & & & & & & & & & & & & & & Xid, Cluster,  \\ 
EROS2C N$^{\circ}$ & Identification      & $\alpha$ \quad & $\delta$  \quad \quad \quad  \quad &  Period      &  $\langle V_{E} \rangle $  &   $\Delta V_{E}$            &               $\langle R_{E} \rangle $  &    $\Delta R_{E}$          &  $R_{21}$   &  $\sigma_{R_{21}}$   &        $R_{31}$ &   $\sigma_{R_{31}}$  &       $\phi_{21}$  &  $\sigma_{\phi_{21}}$        &  $\phi_{31}$   &  $\sigma_{\phi_{31}}$      &   Remarks     \\ 
\hline
\hline
J0052--7341\#5301 & (Cp35k--5117) & 0:52:26.7 & -73:41:05.2 &   5.6632 &  15.59 &  0.92 &  15.12 &  0.66 &   0.467 &  0.016 &  0.189 &  0.014 &  4.424 &  0.043 &  2.462 &  0.091 &      \\ 
J0050--7301\#5302 & (Cp30m--14186) & 0:50:54.0 & -73:01:29.8 &   5.6896 &  15.83 &  0.27 &  15.32 &  0.21 &   0.265 &  0.049 &  0.109 &  0.044 &  4.654 &  0.188 &  3.238 &  0.450 &      \\ 
J0050--7311\#5303 & (Cp30n--16270) & 0:50:08.4 & -73:11:53.3 &   5.7487 &  15.96 &  0.72 &  15.46 &  0.52 &   0.368 &  0.014 &  0.266 &  0.013 &  5.006 &  0.040 &  3.035 &  0.059 & [H86] 112    \\ 
J0043--7410\#5304 & (Cp47k--4565) & 0:43:43.1 & -74:10:46.8 &   5.7498 &  15.34 &  0.89 &  14.96 &  0.63 &   0.426 &  0.020 &  0.246 &  0.017 &  4.858 &  0.050 &  2.754 &  0.087 &      \\ 
J0048--7303\#5305 & (Cp30k--13790) & 0:48:05.1 & -73:03:35.1 &   5.7528 &  15.41 &  0.56 &  15.03 &  0.48 &   0.361 &  0.051 &  0.197 &  0.046 &  4.407 &  0.176 &  2.715 &  0.311 & 1Ot-cepheid ?    \\ 
J0037--7358\#5306 & (Cp44l--5301) & 0:37:29.1 & -73:58:00.3 &   5.7960 &  15.77 &  0.70 &  15.23 &  0.53 &   0.485 &  0.033 &  0.212 &  0.032 &  4.609 &  0.082 &  2.999 &  0.157 &      \\
\hline    
\end{tabular}
\end{flushleft} 
\end{table} 
\clearpage 
\clearpage 
 \begin{table}[pth] 
\tiny 
\begin{flushleft} 
\begin{tabular}{|l@{  }l@{  }l@{  }l@{ }l@{ }r@{ }r@{ }r@{ }r@{ }l@{ }l@{ }l@{ }l@{ }l@{ }l@{ }l@{ }l@{ }l|} 
\hline 
 & & & & & & & & & & & & & & & & & Xid, Cluster,  \\ 
EROS2C N$^{\circ}$ & Identification      & $\alpha$ \quad & $\delta$  \quad \quad \quad  \quad &  Period      &  $\langle V_{E} \rangle $  &   $\Delta V_{E}$            &               $\langle R_{E} \rangle $  &    $\Delta R_{E}$          &  $R_{21}$   &  $\sigma_{R_{21}}$   &        $R_{31}$ &   $\sigma_{R_{31}}$  &       $\phi_{21}$  &  $\sigma_{\phi_{21}}$        &  $\phi_{31}$   &  $\sigma_{\phi_{31}}$      &   Remarks     \\ 
\hline
\hline
J0047--7303\#6001 & (Cp30k--13078) & 0:47:35.6 & -73:03:01.2 &   0.5908 &  18.34 &  0.40 &  17.97 &  0.38 &   0.167 &  0.135 &  0.271 &  0.133 &  3.394 &  0.827 &  0.732 &  0.618 &      \\ 
J0054--7357\#6002 & (Cp35n--3834) & 0:54:19.3 & -73:57:08.9 &   0.5959 &  18.30 &  0.84 &  18.02 &  0.62 &   0.422 &  0.039 &  0.063 &  0.033 &  4.212 &  0.111 &  2.096 &  0.600 &      \\ 
J0039--7314\#6003 & (Cp40n--4383) & 0:39:15.9 & -73:14:26.3 &   0.6505 &  18.53 &  0.43 &  18.22 &  0.41 &   0.086 &  0.062 &  0.171 &  0.063 &  4.328 &  0.745 &  5.511 &  0.413 &      \\ 
J0045--7323\#6004 & (Cp43m--4647) & 0:45:49.4 & -73:23:45.3 &   0.6552 &  17.80 &  0.52 &  17.53 &  0.38 &   0.208 &  0.036 &  0.121 &  0.035 &  3.321 &  0.165 &  1.276 &  0.280 & [H86] 76    \\ 
J0053--7319\#6005 & (Cp33k--7401) & 0:53:10.4 & -73:19:20.8 &   0.6564 &  17.89 &  0.46 &  17.56 &  0.36 &   0.301 &  0.038 &  0.119 &  0.038 &  3.639 &  0.142 &  1.481 &  0.317 &      \\ 
J0048--7309\#6006 & (Cp30l--12636) & 0:48:04.3 & -73:09:55.8 &   0.6717 &  17.88 &  0.41 &  17.49 &  0.36 &   0.260 &  0.068 &  0.187 &  0.072 &  2.982 &  0.307 &  1.531 &  0.434 &      \\
\hline    
\end{tabular}
\end{flushleft} 
\end{table} 
\clearpage 
\clearpage 
 \begin{table}[pth] 
\tiny 
\begin{flushleft} 
\begin{tabular}{|l@{  }l@{  }l@{  }l@{ }l@{ }r@{ }r@{ }r@{ }r@{ }l@{ }l@{ }l@{ }l@{ }l@{ }l@{ }l@{ }l@{ }l|} 
\hline 
 & & & & & & & & & & & & & & & & & Xid, Cluster,  \\ 
EROS2C N$^{\circ}$ & Identification      & $\alpha$ \quad & $\delta$  \quad \quad \quad  \quad &  Period      &  $\langle V_{E} \rangle $  &   $\Delta V_{E}$            &               $\langle R_{E} \rangle $  &    $\Delta R_{E}$          &  $R_{21}$   &  $\sigma_{R_{21}}$   &        $R_{31}$ &   $\sigma_{R_{31}}$  &       $\phi_{21}$  &  $\sigma_{\phi_{21}}$        &  $\phi_{31}$   &  $\sigma_{\phi_{31}}$      &   Remarks     \\ 
\hline
\hline
J0048--7313\#6101 & (Cp30l--18159) & 0:48:57.3 & -73:13:08.1 &   1.1942 &  16.90 &  0.31 &  16.55 &  0.20 &   0.207 &  0.053 &  0.041 &  0.052 &  4.627 &  0.276 &  2.855 &  1.328 &      \\ 
J0053--7323\#6102 & (Cp33k--13370) & 0:53:28.5 & -73:23:25.0 &   1.2012 &  17.26 &  0.48 &  16.84 &  0.34 &   0.292 &  0.035 &  0.087 &  0.037 &  4.317 &  0.129 &  1.834 &  0.352 &      \\ 
J0054--7329\#6103 & (Cp33n--6623) & 0:54:53.6 & -73:29:31.6 &   1.2018 &  17.25 &  0.43 &  16.83 &  0.31 &   0.149 &  0.029 &  0.018 &  0.032 &  4.082 &  0.205 &  2.815 &  1.495 &      \\ 
J0044--7356\#6104 & (Cp45l--3840) & 0:44:17.3 & -73:56:51.6 &   1.2057 &  17.05 &  0.49 &  16.72 &  0.35 &   0.231 &  0.023 &  0.108 &  0.022 &  4.199 &  0.105 &  1.912 &  0.209 &      \\ 
J0052--7342\#6105 & (Cp35k--6474) & 0:52:36.6 & -73:42:29.3 &   1.2062 &  17.22 &  0.34 &  16.81 &  0.26 &   0.161 &  0.037 &  0.045 &  0.037 &  4.835 &  0.239 &  3.560 &  0.776 &      \\ 
J0052--7322\#6106 & (Cp33k--12248) & 0:52:48.3 & -73:22:40.7 &   1.2133 &  17.34 &  0.38 &  16.88 &  0.28 &   0.239 &  0.040 &  0.084 &  0.037 &  4.396 &  0.167 &  1.439 &  0.461 &      \\
\hline    
\end{tabular}
\end{flushleft} 
\end{table} 
\clearpage 
\clearpage 
 \begin{table}[pth] 
\tiny 
\begin{flushleft} 
\begin{tabular}{|l@{  }l@{  }l@{  }l@{ }l@{ }r@{ }r@{ }r@{ }r@{ }l@{ }l@{ }l@{ }l@{ }l@{ }l@{ }l@{ }l@{ }l|} 
\hline 
 & & & & & & & & & & & & & & & & & Xid, Cluster,  \\ 
EROS2C N$^{\circ}$ & Identification      & $\alpha$ \quad & $\delta$  \quad \quad \quad  \quad &  Period      &  $\langle V_{E} \rangle $  &   $\Delta V_{E}$            &               $\langle R_{E} \rangle $  &    $\Delta R_{E}$          &  $R_{21}$   &  $\sigma_{R_{21}}$   &        $R_{31}$ &   $\sigma_{R_{31}}$  &       $\phi_{21}$  &  $\sigma_{\phi_{21}}$        &  $\phi_{31}$   &  $\sigma_{\phi_{31}}$      &   Remarks     \\ 
\hline
\hline
J0054--7317\#6201 & (Cp33m--4200) & 0:54:23.0 & -73:17:58.8 &   2.3821 &  16.48 &  0.18 &  15.95 &  0.14 &   0.131 &  0.068 &  0.030 &  0.068 &  4.983 &  0.544 &  2.885 &  2.265 &      \\ 
J0055--7320\#6202 & (Cp33m--7410) & 0:55:44.3 & -73:20:28.1 &   2.3866 &  16.08 &  0.34 &  15.71 &  0.23 &   0.038 &  0.038 &  0.083 &  0.037 &  1.082 &  0.913 &  4.405 &  0.447 &      \\ 
J0045--7342\#6203 & (Cp43n--4447) & 0:45:04.8 & -73:42:20.6 &   2.4183 &  16.67 &  0.24 &  16.29 &  0.15 &   0.038 &  0.041 &  0.104 &  0.041 &  2.137 &  1.074 &  5.516 &  0.415 &      \\ 
J0052--7348\#6204 & (Cp35l--2791) & 0:52:58.2 & -73:48:02.7 &   2.4187 &  15.91 &  0.19 &  15.50 &  0.16 &   0.077 &  0.074 &  0.049 &  0.073 &  4.155 &  0.968 &  3.085 &  1.539 &      \\ 
J0052--7335\#6205 & (Cp33l--11304) & 0:52:16.1 & -73:35:14.4 &   2.4430 &  15.91 &  0.26 &  15.52 &  0.20 &   0.074 &  0.036 &  0.027 &  0.037 &  3.699 &  0.495 &  5.263 &  1.326 &      \\ 
J0051--7259\#6206 & (Cp30m--11413) & 0:51:09.5 & -72:59:45.5 &   2.4526 &  15.93 &  0.33 &  15.67 &  0.26 &   0.031 &  0.032 &  0.023 &  0.033 &  3.842 &  0.997 &  4.646 &  1.318 &      \\
\hline    
\end{tabular}
\end{flushleft} 
\end{table} 
\clearpage 
\clearpage 

}
\end{document}